\documentclass[a4paper,11pt]{article}
\usepackage{jinstpub} 
\usepackage{lineno}
\usepackage{caption}
\usepackage{subcaption}

\title{Real-time Track Reconstruction with FPGAs: an example from MUonE}

\author[1]{Michael McGinnis \note{Corresponding author.},}
\author{David Monk}
\author{and Kristian Hahn}
\affiliation{Northwestern University,\\Evanston, Illinois, U.S.A}

\emailAdd{michael.mcginnis@cern.ch}

\abstract{The feasibility of real-time track fitting of straight tracks for online event selection is shown in the context of the MUonE experiment. In order to reach the high throughput and low latency required for real-time track fitting, the tracking algorithm is implemented using High-Level Synthesis for an FPGA, which provides provides a balance between flexibility and performance. Hardware tracking results from a test with high-rate muons at the CERN M2 beam are presented. This test, a joint effort by the MUonE and Compact Muon Solenoid (CMS) experiments, operated with Two-Strip (2S) modules provided by the CMS tracker group. The prototype version of the fitter used in this test achieves 3.75\,$\mu$s latency and 7.75\,\% resource use. A comparison of the results with those obtained from a pure software implementation shows agreement well below relevant uncertainty scales. Developments since the test are also presented, with the hardware algorithm now reaching 1.35\,$\mu$s latency and producing pipelined fitting results above 50\,MHz while requiring less resources than the initial version. Doing so required including certain assumptions in the algorithm, but these assumptions hold for linear track fitting in general and algorithm remains useful in contexts other than MUonE.}


\begin{document}
\maketitle
\flushbottom

\section{Introduction}
Charged particle tracking has been an integral aspect of almost every high energy physics experiment\,\cite{Fruhwirth:2020zbo}\cite{Turala:2005dr}. Track reconstruction is the process of collecting fine-grained position measurements, called hits, from tracking detectors and associating them to infer the trajectory (``track'') of a charged particle which may have produced the signals in the detector. Track fitting refers to the process of extracting track parameters from sets of candidate hits. ``Online'' is defined as a method that is performed with hardware as data is received and processed from the detector, before storage. The information gained from online reconstruction of tracks can be used in selecting events for storage and further analysis. A trigger system is sometimes used for event selection, in which the full data is buffered while a subset of the data is used to decide if the event should be kept for further reconstruction and analysis. A multi-level trigger architecture is common, where the first level is often implemented in hardware for high-rate processing, and is limited to a subset of the detector. It then performs some selection and passes events to an upper trigger for further reconstruction and combination with other detector regions. The second stage is often implemented in software and selects which events to send to long-term storage. For instance, the Compact Muon Solenoid (CMS) experiment uses such a trigger architecture \cite{CMS:2008xjf}

Real-time tracking requires high-rate processing, and so is typically implemented in hardware. A field-programmable gate array (FPGA), a reconfigurable integrated circuit, is often used as is provides performance gains via hardware implementation, while also allowing flexibility for development. This requires development with a hardware description language (HDL), but high-level synthesis (HLS) is a tool which allows for C or C++ code to be synthesized into an HDL, allowing for quicker development.

This paper presents an algorithm for real-time tracking of straight tracks for high energy physics experiments. The algorithm uses linear least squares (LLS) regression and is implemented on an FPGA using HLS. The aim was to develop the algorithm to be useful in general, not specific to one experiment. The algorithm has been tested in a beam test with high-rate muons, and further optimized to improve throughput and resource use.

\subsection{Hardware Tracking Examples}
Several experiments have used hardware tracking to mitigate high data rates. Two examples, namely Belle-II, and the CMS upgrade for the HL-LHC, are briefly described here to illustrate the state-of-the-art in hardware tracking.

Belle-II, an experiment studying B physics with electron-positron collisions\,\cite{Skambraks:2020hlt}, uses hardware tracking to reject background events. Much of these background events originate from points displaced along the beamline. Selecting on the $z$ intercept of reconstructed tracks reduces data rates without significant loss in efficiency for events of interest. Hits are generated in a wire chamber composed of 14336 sense wires arranged in 56 cylindrical layers about the beam, and tracks are reconstructed via an FPGA-based system with a 5\,$\mu$s latency budget and a 30\,kHz maximum rate. It has used 2D and 3D Hough transforms since 2019, and a neural network since 2021.

The HL-LHC upgrade\,\cite{ZurbanoFernandez:2020cco} is set to deliver 5 to 7.5 times the nominal LHC luminosity, resulting in approximately 200 proton-proton interactions each time proton bunches cross, as compared to the nominal LHC average of 27. The CMS experiment, one of two general purpose detectors at the LHC, intends to use hardware tracking to improve its event selection in this highly challenging environment.

The CMS experiment will use tracking information in the first level trigger \cite{Holmberg:2024lnq}. It will use data from the outer tracker modules, which each comprise two closely separated sensors. Charged particle trajectories are bent under the influence of a high magnetic field, with higher momentum particles bending less. Correlated hits from the two sensors, called ‘stubs’, are used to only read out those hits which are consistent with tracks having a transverse momentum greater than 2\,GeV. This reduces the data rate sent off the detector by a factor of ten. The stub data enters an FPGA-based tracking implementation with a 4\,$\mu$s latency budget, the fitted information is then used by the Level-1 trigger to further reduce the data rate to the high-level trigger to a maximum of 750\,kHz. The hardware tracking is performed in two steps, first a road-search algorithm\,\cite{tracklet}, then a Kalman Filter\,\cite{kalmanfilter}. Pairs of stubs in consecutive layers seed the algorithm and a beam-spot constraint is used to estimate initial track parameters; then, the seeds are projected to other layers, and stubs within a certain window are associated with the tracks. Once duplicate tracks are removed, the Kalman Filter is used to extract final parameters by iteratively updating the track parameters as each stub is included in the fit.

MUonE \cite{muoneproposal} will also require event selection in order to cope with high data rates. While Belle-II and CMS are triggered experiments, MUonE will perform triggerless readout. This is similar to LHCb, which has triggerless readout in software on GPUs\,\cite{Aaij:2019zbu}. Event selection for MUonE will occur through partially reconstructing each event in real-time and filtering which events are selected for long-term storage. In this manner, data is streamed off detector rather than buffered for a trigger signal. This requires event reconstruction and selection at a rate similar to readout, which is 40\,MHz, as MUonE will use modules and electronics developed for CMS. Thus, FPGAs will be used to benefit from the speed of a hardware implementation, as do Belle-II and CMS. The intent of the MUonE experiment is to measure the angular distribution of elastic scatters between a muon beam and atomic electrons in a fixed target, which has a distinct topology from several main backgrounds, including non-interacting muons and pair production. Here, online tracking should be a powerful tool to select signal events for long-term storage.


\subsection{High-Level Synthesis}
Implementing an algorithm on an FPGA typically requires it to be written in an HDL which is then synthesised into a bitstream to be loaded onto the FPGA. Since writing a complex algorithm like track fitting in an HDL can be difficult and time-consuming, HLS was used to implement the algorithm. Vitis HLS\,\cite{vitishls}, formerly Vivado HLS, is a product from AMD-Xilinx intended to aid in fast development and testing of hardware projects by converting a behavioral description into a register transfer logic (RTL) description. With it, one can synthesise C or C++ code into a packaged intellectual property (IP) block  with an implementation in an HDL. The main steps are to verify the C/C++ accuracy with a user-provided testbench, synthesise the C/C++ into an HDL instance, verify the accuracy of the HDL implementation via cosimulaion of the algorithm and the test bench, and finally package the synthesised code as an IP block. HLS also provides estimates of the performance (such as latency and throughput) and resource use of the implementation. Furthermore, Vitis HLS provides libraries which are optimised for performance; the most useful for the LLS track fitter are the Vitis linear algebra libraries.

HLS enables the rapid development of complicated algorithms by removing concern over  low-level firmware implementation details. Control over how the algorithm is implemented in hardware is traded for the ease of algorithm specification. Some degree of control is however possible through the use of HLS pragmas that can constrain the implementation, such as which resources are used for storage. Pragmas can also be used to enhance performance, prioritising lower latency or lower FPGA resource use. For instance, the extent of loop unrolling,  a process of implementing the loop as each iteration working in parallel rather than as a single process used multiple times, can be used to manage timing and resource constraints. Another important pragma enforces pipelining. Pipelined algorithms continually accept new inputs, whereas a non-pipelined algorithm accepts new inputs only when the previous have been output.

Much of the difficulty of developing with HLS lies in optimizing the algorithm. Broadly, this occurs in two directions, and in general a project requires a combination of both. First, Vitis can be used to decrease latency and increase throughput, for instance by enforcing pipelining or completely unrolling loops. Secondly, a process might be resource limited, so pragmas can be used to assign specific resources or restrict time optimization, for example limiting loop unrolling. Though HLS is used to limit development in an HDL, it does not completely replace it. In the case of using multiple IP blocks, some HDL is needed to connect them. Then, to make best use of HLS, it is important to be aware of the strengths and limitations of hardware implementation.
\section{FPGA-based Real-time Track Fitting}
In order to use track fitting for event selection, the implementation shown here is intended to be ``online'', that is, on hardware within the data acquisition (DAQ) system as data is received and processed from the detector and before long term storage. In most high energy physics experiments, this corresponds to a high data rate, and so the track fitting implementation requires high throughput. An FPGA is used to cope with high rates, while also allowing flexibility for development and testing.

The algorithm shown here assumes that tracks can be constructed as straight lines. This is the case when there are no macroscopic electric or magnetic fields in the tracker, and multiple scattering is assumed to be limited to material outside the tracker modules, such as a target. In this case, the $x$ and $y$ projections of the tracks are independent, and the track fitting may occur separately for each. After the projections are fit, some mechanism is needed to associating the $x$ and $y$ fits to form a 3D track.

\subsection{The Track Fitting Algorithm} \label{sec:algorithm}
The method of LLS is aptly suited to the problem of fitting linear tracks. This can be expressed as:
\begin{equation} \label{lls_prob}
    \boldsymbol{m} = \boldsymbol{F \pi} + \boldsymbol{\epsilon}
\end{equation}
where \textbf{\textit{m}} is an $n$-vector of measurements, \textit{x} or \textit{y} coordinates of hits; \textbf{\textit{F}} is the Jacobian, an $n$-by-two matrix formed from the \textit{z} coordinates of hits in the first column and ones in the second; $\boldsymbol{\pi}$ is a vector of the track parameters, the slope and intercept; and $\boldsymbol{\epsilon}$ is an $n$-vector of unknown measurement errors with a zero expectation value and an $n \times n$ covariance matrix \(\textbf{\textit{G}}^{-1}\) \,\cite{Fruhwirth:2020zbo}. The goal of track fitting is to calculate $\boldsymbol{\pi}$. However, \textbf{\textit{F}} is generally non-invertible, as in this case where there are more hits than parameters. As the problem is over-constrained and influenced by a priori unknown measurement errors, $\boldsymbol{\pi}$ must be estimated.

To estimate the track parameters, \textit{\textbf{F}} and \textit{\textbf{m}} are first scaled by the square root of \textit{\textbf{G}}, \(\textbf{\textit{G}}_{sqrt}\). Then, QR factorization (QRF)\,\cite{LinAlg} is used to factor the scaled \textit{\textbf{F}}, \(\textbf{\textit{F}}_{sc}\), into a product of matrices, \textbf{\textit{QR}}, where \textbf{\textit{Q}} is orthogonal and \textbf{\textit{R}} is right triangular. These are invertible, so that $\boldsymbol{\tilde{\pi}}$, the estimator of $\boldsymbol{\pi}$, can be calculated as:
\begin{equation}
    \boldsymbol{\tilde{\pi}} = \boldsymbol{R^{-1}} \boldsymbol{Q^{-1}} \boldsymbol{m_{sc}}
\end{equation}

and because the inverse of an orthogonal matrix is its transpose:  
\begin{equation}
    \boldsymbol{\tilde{\pi}} = \boldsymbol{R^{-1}} \boldsymbol{Q^{T}} \boldsymbol{m_{sc}} .
    \label{eqn:params}
\end{equation}

Meanwhile, the fit parameter covariances are calculated by\,\cite{Fruhwirth:2020zbo}:
\begin{equation}
    cov( \boldsymbol{\tilde{\pi}} ) = (\boldsymbol{F}^{T} \boldsymbol{G F})^{-1} .
    \label{eqn:cov}
\end{equation}

The above assumes that measurements have been transformed to a consistent, global Cartesian coordinate system, but is independent of the tracker geometry. The size of the inputs, \textbf{\textit{m}} and \textbf{\textit{F}}, are fixed for a given implementation, but they can scale to accommodate different numbers of hits. However, the size will impact the resource use and performance of the implementation.



\subsection{Implementation in High Level Synthesis}
\begin{figure}[ht]
    \centering
    \includegraphics[width=0.6\textwidth,keepaspectratio=true]{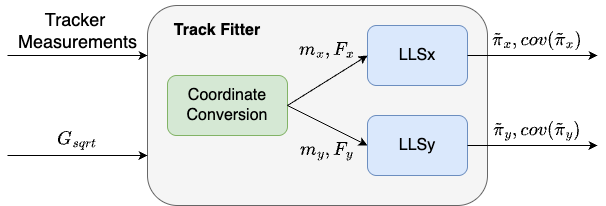}
    \caption[Fitter Schematic 1]{A block diagram of the track fitter showing the coordinate conversion block in green, and the separate LLS fitting blocks in blue. The inputs to the coordinate conversion block are read out from the sensors, and the measurement uncertainty is input to the LLS blocks. The outputs are the track parameter and parameter covariances for each $x$ and $y$. The coordinate conversion block is specific to the tracker geometry and can be adjusted for a given case.}
    \label{fig:fitterschematic1}
\end{figure}

\begin{figure}[ht]
    \centering
    \includegraphics[width=0.6\textwidth,keepaspectratio=true]{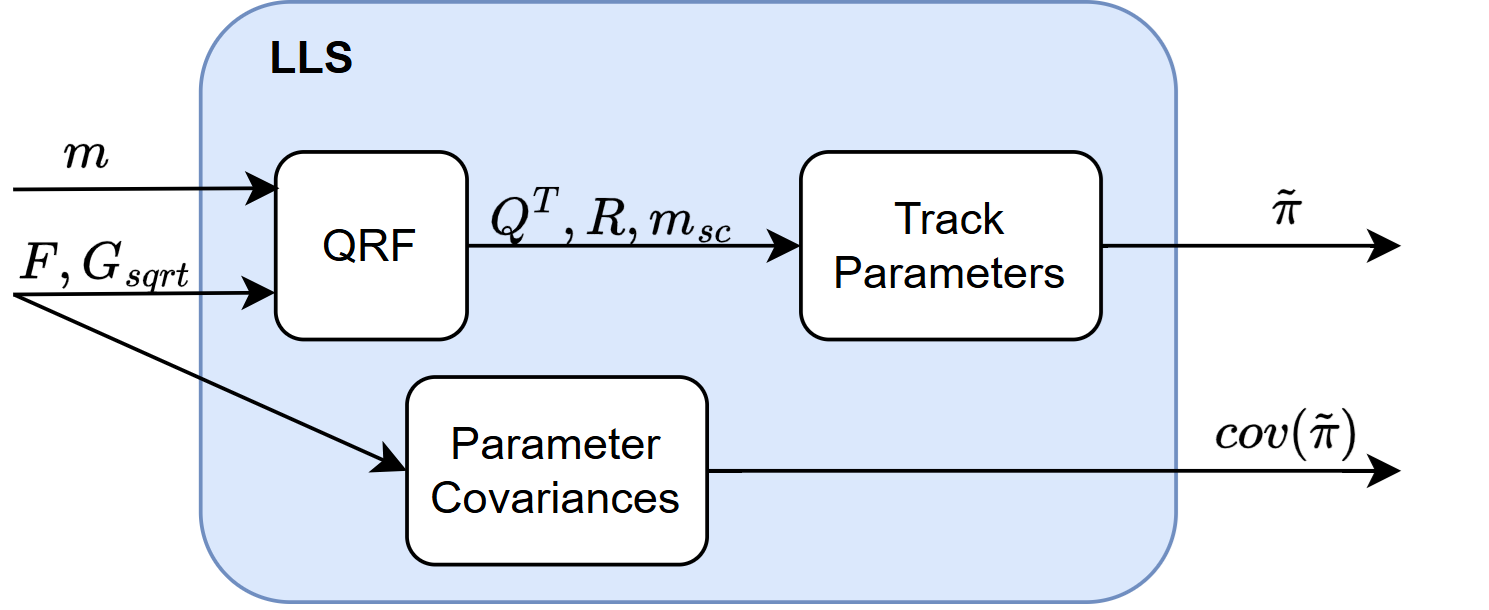}
    \caption[Fitter Schematic 2]{A block diagram of an LLS block from the track fitter. It is composed of a block performing QRF on \(\textbf{\textit{F}}_{sc}\), and blocks to calculate the outputs, the track parameters, and parameter covariances. \(\textbf{\textit{F}}\) and \(\textbf{\textit{G}}_{sqrt}\) are inputs to both the QRF and covariance blocks, while \(\textbf{\textit{m}}\) is an additional input to QRF, which outputs the results of QR decomposition and \(\textbf{\textit{m}}\) scaled by \(\textbf{\textit{G}}_{sqrt}\) to the block calculating the track parameters.}
    \label{fig:fitterschematic2}
\end{figure}

The algorithm in the previous section was implemented using HLS for an FPGA, and figures \ref{fig:fitterschematic1} and \ref{fig:fitterschematic2} depict the implementation. In particular, figure \ref{fig:fitterschematic1} shows the complete algorithm architecture, including a global coordinate conversion block (green, explained below), and an identical LLS solver for each $x$ and $y$ projection (blue). Each LLS solver performs the calculations shown in section \ref{sec:algorithm}. Figure \ref{fig:fitterschematic2} shows the details of a single LLS solver.

In this scheme, one global coordinate conversion block sends its outputs to two LLS fitters, one to fit in $x$ and the other in $y$. Its inputs are the outputs from the sensors which describe the hits in the sensor coordinates. These are converted to global coordinates, and in general should include corrections calculated by performing an alignment procedure on previous data. The global conversion block also outputs the Jacobians (\textbf{\textit{F}}), composed of the $z$ coordinates of the hits, for each fit. It is important to note that the global coordinate conversion block is the only experiment-specific portion of the algorithm. In principle, the algorithm can be extended to any straight-line tracking with appropriate changes to this block.

The LLS solver is composed of a block which scales \textbf{\textit{F}} and performs the QRF of \(\textbf{\textit{F}}_{sc}\), a block which uses the factors to calculate the track parameters (equation \ref{eqn:params}), and a block to calculate the parameter covariances (equation \ref{eqn:cov}). Two copies of the LLS solver are used to perform the fits in $x$ and $y$ and generate 3D tracks. The QRF block takes \textbf{\textit{m}}, \textbf{\textit{F}}, and \(\textbf{\textit{G}}_{sqrt}\), outputting \(\textbf{\textit{m}}_{sc}\), and the results of QRF of \(\textbf{\textit{F}}_{sc}\), namely \textbf{\textit{R}} and the transpose of \textbf{\textit{Q}}. The parameter block then takes \(\textbf{\textit{Q}}^{T}\), \textbf{\textit{R}}, and \(\textbf{\textit{m}}_{sc}\) as inputs, calculates the parameters as in equation \ref{eqn:params}, and outputs them. The covariance block takes \textbf{\textit{F}} and \(\textbf{\textit{G}}_{sqrt}\), then outputs the covariances. As shown in the figure \ref{fig:fitterschematic2}, the covariance block is independent of the other two, and so runs in parallel as is possible with hardware implementations.
\section{2023 Test Beam}
In the autumn of 2023, a reduced version of MUonE was tested, providing a proof of concept for the experiment as well as verifying the module and DAQ performance. This was also an opportunity to test real-time track fitting with an FPGA. The 2023 test was conducted at the M2 beam at CERN that provides a high intensity muon beam with a maximum rate of $5\times10^7\,\mu/s$.

\subsection{Tracking System for MUonE}
\begin{figure}[ht]
    \centering
    \includegraphics[width=0.5\textwidth,keepaspectratio=true]{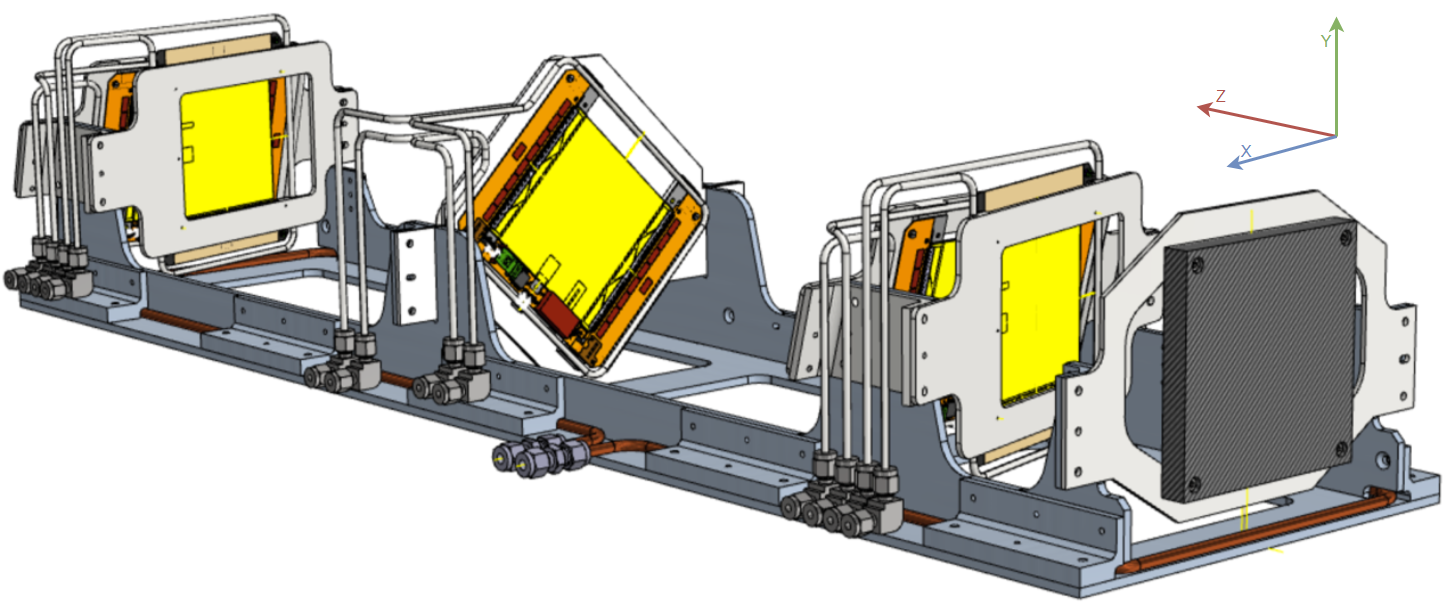}
    \caption[MUonE Station]{CAD drawing of a single station, with tracking sensors in yellow and target in grey in the foreground. The coordinates are shown as well, with the $z$ axis along the beam. The modules are arranged in pairs to measure along perpendicular axes, and the center pair are rotated by 45{\textdegree} so that fits along $x$ and $y$ axes can be combined to form 3D tracks.}
    \label{fig:station_with_axis}
\end{figure}

The MUonE tracker is made of a series of stations, and each station comprises six strip tracking modules which only measure precisely along one axis. One station is is illustrated in figure\,\ref{fig:station_with_axis}.
A single station includes a scattering target and three pairs of perpendicularly arranged modules as: $xy$, $uv$, $xy$. The modules are arranged in perpendicular pairs so that a precise measurement in both axes can be made at a similar position in $z$. The $uv$ pair is rotated by 45 degrees relative to the other pairs, to associate hits along $x$ and $y$ axes when there are multiple possible combinations of hits of $xy$ pairs. For $x$ and $y$ fits originating from a single particle will share hits in the $uv$ modules.

Tracking will be performed using Two-Strip (2S) modules\,\cite{CERN-LHCC-2017-009} developed for the CMS outer tracker for the High-Luminosity upgrade of the LHC. These are comprised by two silicon sensors, each with two columns of 1016 90\,$\mu$m-pitch strips. The strips are 5\,cm long, and have a total active area of $10\times10$\,cm$^2$. Charge sharing between strips allows for the module to resolve hits at 45\,$\mu$m, or half a strip, by assigning the midpoint of two adjacent firing strips to be the center of the hit. The sensors are read out at 40\,MHz, corresponding to the LHC clock.

\subsection{Data Acquisition System}
The DAQ system used to read out data from the tracker is also derived from a proposed system to be used by CMS, namely the Serenity Platform\,\cite{Rose2018Serenity:Phase-2} and Extensible, Modular data Processor (EMP) firmware framework.
Serenity is an ATCA-compatible processing system, capable of housing up to two AMD-Xilinx Ultrascale+\,\cite{AMDXilinx2022UltraScaleOverview} FPGAs.


Each tracking module is connected optically to the Serenity card to transfer the hit data into the FPGA, with four 10\,Gbps Ethernet links connected to transfer data to PCs for consolidation and chunking. Long term data storage is handled by the EOS\,\cite{Peters2015EOSCERN} cluster at CERN. Finally, the stub data is reconstructed offline to calculate the outgoing scattering angles.

The test beam included two stations with twelve 2S modules connected to a single Serenity card, with a 2\,cm carbon target mounted between the stations.
With this setup, it is possible to track the incoming muons in the upstream station and the scattering products in the downstream one. Online track fitting was implemented in parallel with the mainline DAQ, as described in the following section. 


\subsection{Integration with Mainline DAQ}
For the 2023 MUonE test beam, the track fitter was implemented parallel to the mainline DAQ. The mainline DAQ ran without event selection, recording all the test run data, while the track fitter ran on a subset of the data and track parameters were read out through RAMs to a PC.



Track fitting was only implemented in the first station, so that tracks resulted from single muons rather than scattering events. Not all events were selected to be fit, primarily because of the latency of the track fitting algorithm. To reduce the input rate, two cuts were made: the first selecting in time, and second in hit multiplicity. The modules read out eight bunch crossings at a time, so for simplicity and to reduce the input rate to the track fitter, only the first of the bunch crossings were potentially fit. Secondly, events were selected if they had exactly one hit in each module, and so no track finding was necessary as there was only a single possible candidate. A firmware block to combine two dimensional tracks into three dimensional ones was also unneeded as only events with a single track candidate were considered.

In the future, it will be necessary to include track finding to select track candidates to be fit, in order to process events with more than one track. Vertexing must also be included for experiments which aim to measure a scattering process.
\section{Results}
\begin{figure}[ht]
    \centering
    \begin{subfigure}[t]{0.4\textwidth}
        \centering
        \includegraphics[width=\textwidth,keepaspectratio=true] {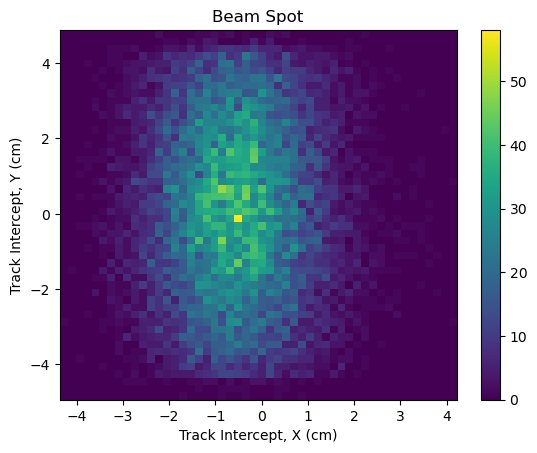}
        \label{fig:onlineintercepts}
    \end{subfigure}
    \begin{subfigure}[t]{0.4\textwidth}
        \centering
        \includegraphics[width=\textwidth,keepaspectratio=true] {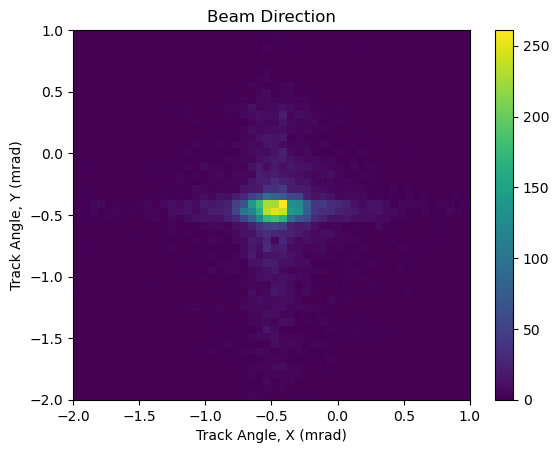}
        \label{fig:onlineangles}
    \end{subfigure}
    \caption[Beam Spots and Directions]{Plots of the beam spot (left) and beam direction (right), found by plotting the intercepts and slopes of the tracks, respectively. The track parameters result from online tracking tested with beam in 2023.}
    \label{fig:beamspotdirection}
\end{figure}

The track fitter implementation described above was tested live in 2023. By plotting the track parameters, it is possible to visualize the beam spot from the $x$ and $y$ intercepts, and the beam direction from the slopes. Figure \ref{fig:beamspotdirection} shows these plots with track data produced by the online fits. Both the measured beam spot position and the beam direction are consistent with offline reconstruction of the same parameters.

\begin{figure}[ht]
    \centering
    \includegraphics[width=0.5\textwidth,keepaspectratio=true]{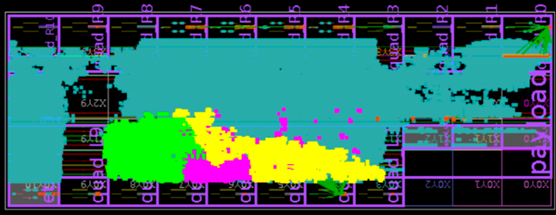}
    \caption[FPGA Resource Use]{The layout of the MUonE DAQ hardware usage on a Xilinx KU15P FPGA for the 2023 Fall Test Beam at the CERN M2 beamline. The yellow and green portions correspond to the fitters in each dimension (6.7\,\% of the total resources), pink to the global coordinate converter and other fitter infrastructure (1.04\,\%), and blue to the mainline DAQ.}
    \label{fig:fpgaresourceuse}
\end{figure}

The track fitting algorithm had a latency of approximately 1200 cycles with a 320\,MHz clock, equivalent to 3.75\,$\mu$s. The average period of single track events were estimated with test beam data to be 2.5\,$\mu$s. Because only one bunch crossing of eight were fit, the throughput of the prototype was sufficient. The initiation interval (II), the number of cycles until the algorithm can process another set of inputs, is equal to the latency, as no pipelining was implemented in the prototype fitter. The track fitter required 7.75\,\% of the resources of a Xilinx KU15P FPGA, with the global coordinate converter being 0.91\,\%, and each LLS solver requiring an average of 3.35\,\%. Figure \ref{fig:fpgaresourceuse} shows the layout of the DAQ on the FPGA, with yellow and green corresponding to the fitters in each dimension, pink to the global coordinate converter and other fitter infrastructure, and blue the mainline DAQ.

\begin{figure}[ht]
    \centering
    \begin{subfigure}[t]{0.4\textwidth}
        \centering
        \includegraphics[width=\textwidth,keepaspectratio=true] {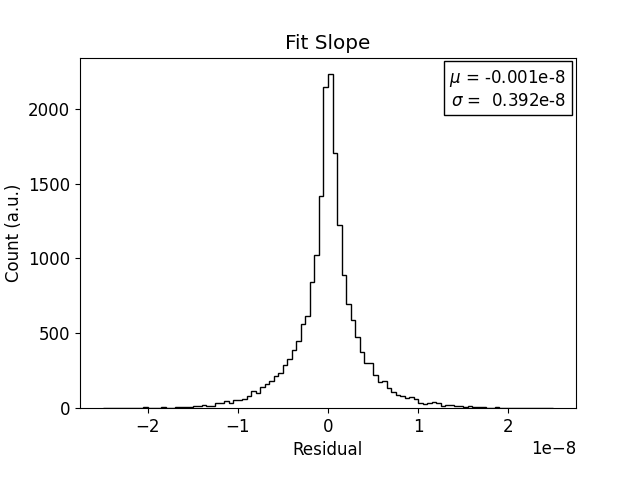}
        \label{fig:Slope}
    \end{subfigure}
    \begin{subfigure}[t]{0.4\textwidth}
        \centering
        \includegraphics[width=\textwidth,keepaspectratio=true] {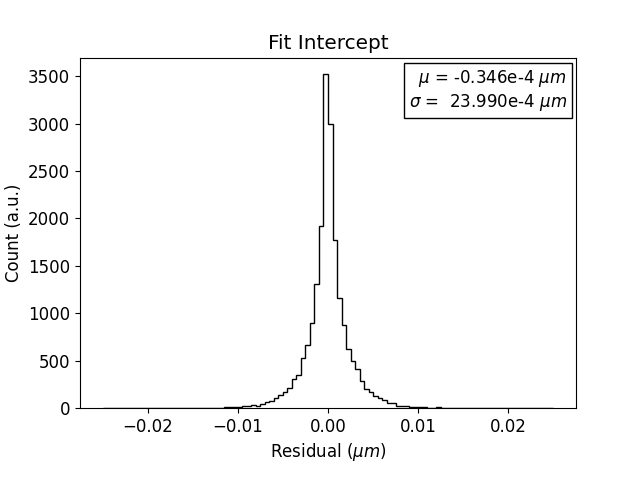}
        \label{fig:Intercept}
    \end{subfigure}
    
    \begin{subfigure}[t]{0.3\textwidth}
        \centering
        \includegraphics[width=\textwidth,keepaspectratio=true] {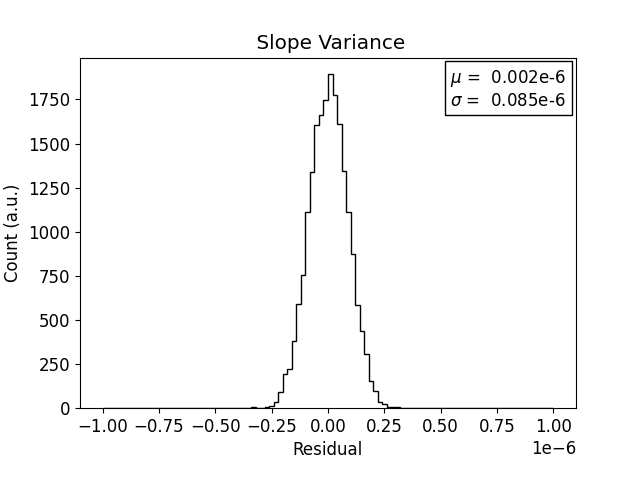}
        \label{fig:covP_00}
    \end{subfigure}
    \hfill
    \begin{subfigure}[t]{0.3\textwidth}
        \centering
        \includegraphics[width=\textwidth,keepaspectratio=true] {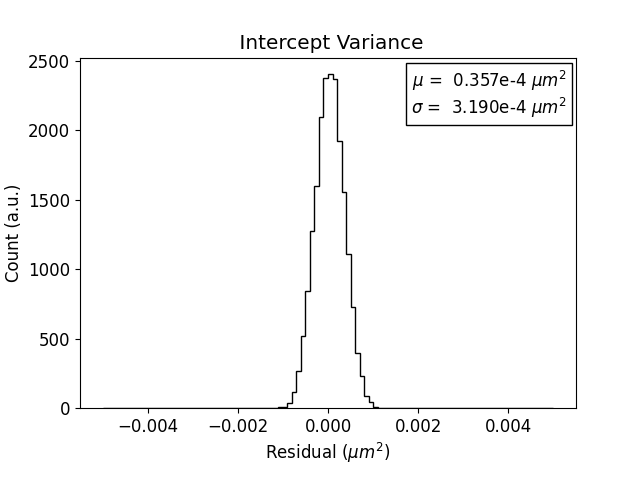}
        \label{fig: covP_11}
    \end{subfigure}
    \hfill
    \begin{subfigure}[t]{0.3\textwidth}
        \centering
        \includegraphics[width=\textwidth,keepaspectratio=true] {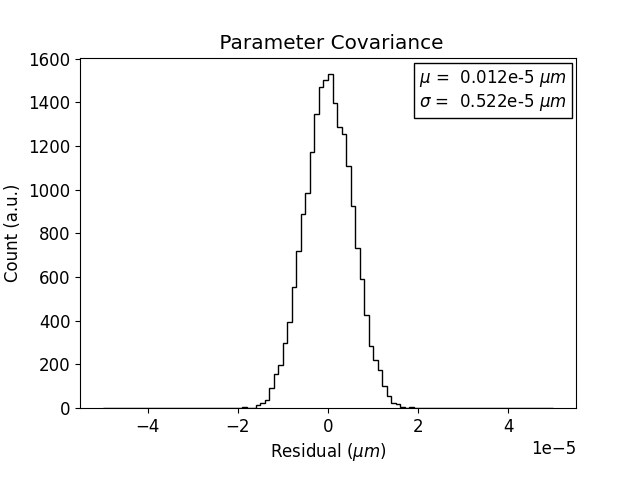}
        \label{fig: covP_01 }
    \end{subfigure}
    \caption[Online Offline Comparison]{Plots showing the agreement between online and offline track reconstruction using data from the 2023 MUonE test beam. The results are combined for the X and Y fits as they are independent, and all outputs from the online fitter are shown: track parameters and track covariances.}
    \label{fig:onlineofflinecomp}
\end{figure}

In addition to verification with the Vitis cosimulation, the hardware implementation was verified by manually loading hit data to the firmware on an FPGA via RAMs. Fit results using these input hits are then compared to the results from a software fit using the same inputs, to confirm that that the hardware and software outputs are consistent. This was done for ten thousand fits using data collected during the 2023 test run, shown in figure \ref{fig:onlineofflinecomp}. More specifically, these plots show the residuals between the track parameters and covariances found via the offline and online procedures. The distributions are centered close to zero, meaning the hardware implementation is free of bias, and the standard deviations are smaller than the relevant uncertainty scales, showing good agreement between the reconstruction methods. For example, the standard deviation for the fit intercept distribution ($24 \times 10^{-4}\,\mu$m) is four orders of magnitude smaller than the measurement uncertainty due to the strip pitch, 26\,$\mu$m. Also, it is likely that non-zero residuals are due to floating point precision differences, rather than a bias in the implementations.

\subsection{Subsequent Developments and Outlook}
Since the test beam in fall 2023, two improvements to the implementation have been explored: pipelining of the algorithm for higher throughput and custom versions of HLS linear algebra library functions. The protoype implementation that was tested live in 2023 was not pipelined, so that the fitter accepted new inputs only after completing a fit. Pipelining allows algorithm to continuously accept new inputs, resulting in a lower II and a higher rate of fit tracks. Pipelining itself is not intended to change the algorithmic latency substantially, but the changes required by the synthesis tool to implement pipelining was observed to decrease overall latency. To minimize the II, calculations were organized to limit data dependencies and structure data to flow from one step to the next as much as possible.

The second main improvement resulted from the custom implementation of linear algebra functions, whereas the prototype implementation relied on the Vitis HLS linear algebra libraries. These functions include QR Factorisation and matrix multiplication. Generally, these improvements exploited the simplicity of the inputs, including their traits into the algorithm as assumptions. In contrast, the library functions had no assumptions on the inputs other than the matrix sizes. For example, since \textbf{\textit{R}} is right-triangular and \(\textbf{\textit{G}}_{sqrt}\) is diagonal, some entries are zero, while those entries need not be calculated, and matrix operations with \textbf{\textit{R}} and \(\textbf{\textit{G}}_{sqrt}\) are greatly simplified. Similarly, the parameter covariances are symmetric, reducing the number of elements which need to be calculated.

Another important assumption is that every entry of the second column of \textit{\textbf{F}} is always one, and the second column of \(\textbf{\textit{F}}_{sc}\) is the diagonal entries of \(\textbf{\textit{G}}_{sqrt}\). So, multiplication of the second column of \textit{\textbf{F}} by a row vector is simply addition. More importantly, this is used to simplify the QRF process. This was implemented through the Gram-Schmidt procedure\,\cite{LinAlg}, in which \textbf{\textit{Q}} is composed of independent vectors formed from the columns of \(\textbf{\textit{F}}_{sc}\). The second column of \(\textbf{\textit{F}}_{sc}\) is then chosen as the first basis vector, and some calculations can be done once by hand and hard-coded into the algorithm. So, the required calculation to find \textbf{\textit{Q}} is normalising the second column of \(\textbf{\textit{F}}_{sc}\), then finding the component of first column of \(\textbf{\textit{F}}_{sc}\) which is perpendicular to the second column, and normalizing it.

The assumptions outlined above are general to linear track fitting via QRF. That \textbf{\textit{R}} is right-triangular and the parameter covariance matrix is symmetric are true by definition. \(\textbf{\textit{G}}_{sqrt}\) is diagonal under the assumption that the measurements, and therefore the measurement errors, are independent. Finally, the second column \textbf{\textit{F}} is all ones because it multiplies the second entry of $\boldsymbol{\pi}$, the intercept, to yield a linear equation. This can be seen by considering equation \ref{lls_prob}. Or, since \textbf{\textit{F}} is the Jacobian of a function parameterised by a slope and intercept, the elements of the second column are  derivatives of linear functions with respect to the intercepts, which is one.

Additionally, by inspecting the algebraic expressions for the matrix multiplications element-by-element, common objects could be found, calculated a single time, and used in multiple other calculations. For example, many computations required the sum of the $z$ coordinates of hits, or the sum of the $z$ coordinates squared. 

Table \ref{Table:algo_perf} summarises the performance and resource use attained with these optimisations, estimated using Vitis HLS. Performance metrics for three optimisation schemes are compared with those of the prototype implementation, where version A applies pipelining to the prototype version, version B uses custom functions, and version C applies pipelining to version B. Fit results from these optimised implementations were again compared with those from software, and no significant differences were observed. The table includes latency and II to characterise throughput, and four representative FPGA resources. A BRAM is used to store large amounts data, up to the scale of kilobits, whereas flip-flops store individual bits. A DSP performs complicated calculations or operations on a large number of bits and a LUT performs simple operations analogous to a series of logic gates.

\begin{table}[ht!]
\begin{center}
\begin{tabular}{||c | c  c  c  c||} 
 \hline
   & Prototype & A & B & C \\ [0.5ex] 
 \hline\hline
 Latency (cycles [$\mu$s]) & 1369 [4.28]  & 852 [2.66] & 701 [2.19] & 431 [1.35] \\ 
 \hline
 II (cycles [$\mu$s]) & 1372 [4.28] & 470 [1.47] & 704 [2.20] & 6 [0.02]\\
 \hline
 LUTs (percent) & 7.74 & 10.82 & 4.78 & 4.57 \\
 \hline
 Flip-Flops (percent) & 7.02 & 10.43 & 4.52 & 4.42 \\
 \hline
 DSP (percent) & 9.30 & 22.41 & 6.55 & 6.66 \\
 \hline
 BRAM (percent) & 4.47 & 3.86 & 2.24 & 2.24 \\[1ex] 
 \hline
\end{tabular}
\caption{Performance and resource use for each fitter version. The time characteristics are estimated by Vitis HLS, while the resource use is found through Vivado implementation. The prototype version was tested in 2023 and relies on the HLS linear algebra libraries. Version A is the prototype version with pipelining applied, while version B uses custom linear algebra implementations, and version C has pipelining applied to version B.}
\label{Table:algo_perf}
\end{center}
\end{table}

Results from version C are most notable, as with an II of 6 cycles and a 320\,MHz clock, single fit results are produced at 53.33\,MHz. This important in the context of tests at the CERN M2 beam, as the results are calculated above the peak muon incidence rate from the CERN M2 beam ($5\times10^7\,\mu/s$). However, a single fitter instance will not be sufficient to fit every event. When scattering events or noise occur, there will be multiple track candidates, requiring more track fitter instances. Notably, this improvement required the custom implementation of the linear algebra functions, as the Vitis library functions could only be pipelined to an II of 470 cycles (shown in column A).

It is generally expected that pipelining an algorithm will decrease the II, while requiring more resources to do so. This occurs as the synthesis tool uses more resources to perform calculations or increases throughput by duplicating operations, for instance by implementing each loop iteration in parallel rather than sequentially. This can be seen when comparing the Prototype and A columns, as the II reduced significantly from 1372 cycles to 470 cycles, and most of the resource uses increased. As noted previously, the latency was much lower due to changes required implement pipelining. In contrast to the other resources, the BRAM use decreased, likely due to the removal of intermediate stages within an IP block due to data streaming between operations.

Comparing the Prototype column to column B shows the benefit from the custom linear algebra implementations. The latency (and non-pipelined II) is almost halved from 1369 cycles to 701 cycles, and the resource use is much lower. These decreases correspond to the algorithm being simplified by the use of custom functions. This simplicity also allows for the improvements when pipelining this implementation, as seen in columns B and C. As expected, the latency and II are reduced, but notably the II for column C is now 6 cycles, compared to the 470 cycles showed in column A. The BRAM use does not change, perhaps reflecting fewer intermediate stages in the non-pipelined algorithm. Of the other resources, only DSPs increase, while LUTs and flip-flops decreased. This is potentially caused by having few opportunities to replicate operations performed by LUTs and flip-flops through means such as loop unrolling, and that some processes previously using those resources were optimized to use DSPs.

Further improvements may be made to the algorithm to improve throughput, in particular by decreasing the II. These improvements mainly concern the input and output of data from the IP blocks. For instance, the current implementations use a single RAM per variable to transfer data between IP blocks, requiring multiple cycles in most cases. By partitioning the variables, it will be possible to use fewer cycles at the cost increasing resource use. Another improvement would be to combine the QRF and Track Parameters blocks, so that transferring data between those IP blocks is not needed. This would decrease both II and resource use. In the current design, the blocks are separate for flexibility when placing them on an FPGA, as the synthesis tools manage resource layout. With the significant decrease in resource use for versions B and C, the combined block may be small enough to be sufficiently flexible for placement.
\section{Conclusions}

For many particle physics experiments, real-time event selection is crucial for reducing bandwidth requirements for data transfer to long-term storage. FPGA-based track reconstruction provides a powerful means of achieving this; data products from these computations may be used to preferentially store events with signal-like characteristics. We have demonstrated that an FPGA implementation can provide the high throughput required for streaming reconstruction while keeping algorithmic latency low. We exercised a first prototype of our hardware real-time track fitter in a 2023 test beam and validated its results against a pure offline software implementation. To our knowledge, this work represents the first example of HLS-based tracking used in data-taking. Improved versions of the algorithm have been shown to perform with tracking results provided above 50\,MHz. This required custom implementations of linear algebra functions, including assumptions on the inputs which are general to linear track fitting. Further improvements and extensions to the algorithm are in development, including improving track fitting performance and track finding and vertexing.

\acknowledgments

This material is based upon work supported by the National Science Foundation under Grant No. 2111556. We acknowledge the MUonE collaboration, their efforts in the test beam, and the opportunity to test the real-time reconstruction. We also acknowledge the contribution of the CMS Tracker group who provided the modules used to collect data and the DAQ hardware and firmware framework, as well as contributed to the joint test beam efforts. We specifically acknowledge the work of Martin Delcourt, Matteo Magherini, and Mark Pesaresi.


\bibliographystyle{JHEP}
\bibliography{biblio.bib}

\end{document}